\documentclass[twocolumn]{aastex631}

\usepackage{bm}
\usepackage{amsmath,amsthm,amsfonts,amssymb,amscd}
\received{XXX X, XXXX}
\revised{XXX X, XXXX}
\accepted{XXX X, XXXX}

\submitjournal{ApJ}

\shorttitle{PAC. I. methods}
\shortauthors{XU et al.}

\begin{document}

\title{Photometric objects Around Cosmic webs (PAC) delineated in a spectroscopic survey. I. Methods}

\correspondingauthor{Yipeng Jing}
\email{ypjing@sjtu.edu.cn}

\author[0000-0002-7697-3306]{Kun Xu}
\affil{Department of Astronomy, School of Physics and Astronomy, Shanghai Jiao Tong University, Shanghai, 200240, People’s Republic of China}

\author{Yun Zheng}
\affil{Department of Astronomy, School of Physics and Astronomy, Shanghai Jiao Tong University, Shanghai, 200240, People’s Republic of China}

\author[0000-0002-4534-3125]{Yipeng Jing}
\affil{Department of Astronomy, School of Physics and Astronomy, Shanghai Jiao Tong University, Shanghai, 200240, People’s Republic of China}
\affil{Tsung-Dao Lee Institute, and Shanghai Key Laboratory for Particle Physics and Cosmology, Shanghai Jiao Tong University, Shanghai, 200240, People’s Republic of China}



\begin{abstract}
 We provide a method for estimating the projected density distribution $\bar{n}_2w_p(r_p)$ of photometric objects around spectroscopic objects in a spectroscopic survey.  This quantity describes the distribution of Photometric sources with certain physical properties (e.g. luminosity, mass, color etc) Around Cosmic webs (PAC) traced by the spectroscopic objects. The method can make full use of current and future deep and wide photometric surveys to explore the formation of galaxies up to medium redshift ($z_s < 2$)\footnote{Throughout the paper, we use $z_s$ for redshift, $z$ for the z-band magnitude, $z_{p}$ for the photometric redshift}, with the aid of cosmological spectroscopic surveys that sample only a fairly limited species of objects (e.g. Emission Line Galaxies).   As an example, we apply the PAC method to the CMASS spectroscopic and HSC-SSP PDR2 photometric samples to explore the distribution of galaxies for a wide range of stellar mass from $10^{9.0}{\rm M_\odot}$ to $10^{12.0}{\rm M_\odot}$ around massive ones at $z_s\approx 0.6$. Using the abundance matching method, we model $\bar{n}_2w_p(r_p)$ in N-body simulation using MCMC sampling, and accurately measure the stellar-halo mass relation (SHMR) and stellar mass function (SMF) for the whole mass range.  We can also measure the conditional stellar mass function (CSMF) of satellites for central galaxies of different mass. The PAC method has many potential applications for studying the evolution of galaxies. 
 \end{abstract}

\keywords{Galaxy formation(595) --- Observational cosmology(1146) --- Galaxy dark matter halos (1880)}


\section{Introduction} \label{sec:intro}
Modeling galaxy formation in the cosmological context is one of the greatest challenges in astrophysics and cosmology today. In past few decades, the broad contours of galaxy formation physics are investigated and built \citep{2010gfe..book.....M,2012RAA....12..917S,2015ARA&A..53...51S}. In theory, cosmic structures grow by gravitational instability from the initial tiny quantum fluctuations generated during the inflationary epoch. Dark matter (DM) halos, which are defined as dark matter objects with the density hundreds times the density of the Universe, are formed around the density peaks of the Universe. Halos grow by accreting surrounding smaller DM halos and diffuse matter including DM and gas \citep{2009ApJ...707..354Z}. The gas is heated by shocks when accreted into a halo \citep{2006MNRAS.368....2D}. Through radiation, the hot gas loses its energy and cools, and the cold gas spirals into the halo center to form a galaxy, which is typically a spiral. Under the hierarchical growth of the DM halo, surrounding galaxies can also be accreted into the halo and become its satellites. Some satellite galaxies, especially relatively massive ones, spiral into the center and coalesce with the central galaxy due to dynamical friction \citep{1994MNRAS.271..676L,2008ApJ...675.1095J}. A merger of a central disk galaxy with a satellite of comparable mass (say, a mass ratio $\geq 0.2$) may significantly change its internal structure, either producing an elliptical galaxy or a disk galaxy with a significant bulge \citep{2006ApJ...636L..81N, 2006ApJ...650..791C, 2011MNRAS.415.1783B}. Studies find that black hole (BH) mass is highly correlated with bulge mass \citep{2013ARA&A..51..511K}, so we expect there are supermassive black holes (SMBHs) within elliptical galaxies or bulges. The strong energy output and/or material outflow produced by an SMBH can heat and/or blow out the surrounding cold gas, thus suppressing the gas reservoir from which stars form and making the host galaxy look red \citep{2012ARA&A..50..455F}. Therefore, the bimodal distribution of galaxy color is highly correlated with the galaxy morphology distribution, with ellipticals being red and spirals being blue.

To fully understand the galaxy properties and distributions, one needs to consider all the complicated physical process involved in galaxy formation and evolution mentioned above. Although there exists  a broad contour of the galaxy formation process, a fully predictive framework from first principles has yet to be established \citep{2017ARA&A..55...59N}. Physical models such as semi-analytic models \citep{1991ApJ...379...52W,1993MNRAS.264..201K,2005ApJ...631...21K,2013ApJ...767..122G} and hydrodynamic simulations \citep{2015MNRAS.446..521S,2018MNRAS.473.4077P,2019MNRAS.486.2827D} approximate physics below their respective resolution scales to simulate the effects of supernovae, radiation pressure, multi-phase gas, black hole accretion, active galactic nucleus (AGN) feedback and metallicity evolution in galaxy formation. However, different approximations lead to different galaxy properties \citep{2008MNRAS.391..481S,2014ApJ...795..123L} and the uncertainty still remains.

Empirical modeling such as halo occupation distribution (HOD, \citet{1998ApJ...494....1J,2000MNRAS.318.1144P,2000ApJ...543..503M,2000MNRAS.318..203S,2002ApJ...575..587B,2003MNRAS.339.1057Y,2005ApJ...633..791Z,2015MNRAS.454.1161Z}) and abundance matching (AM, \citet{1998ApJ...506...19W,2010MNRAS.402.1796W,2013MNRAS.428.3121M,2019MNRAS.488.3143B}) uses significantly weaker prior and the physical constraints come almost entirely from observations. These models connect the average galaxy properties such as occupation numbers and stellar mass to halos as a function of halo mass or halo circular velocity, and determine the parameters by fitting the observed properties of galaxies. Although more complex models have been attempted recently to incorporate properties of gas \citep{2015MNRAS.449..477P}, metallicity \citep{2016MNRAS.462..893R} and dust \citep{2018ApJ...854...36I} to compare with observations, the stellar-halo mass relation (SHMR) is still one of the most commonly used relations to model galaxy-halo connection \citep{2010MNRAS.404.1111G,2010MNRAS.402.1796W,2013MNRAS.428.3121M}, in which larger halos host larger galaxies with a relatively tight scatter. Recent studies found that galaxies with different properties, such as color, may have different SHMR, which may indicate the existence of so called galaxy assembly bias that causes the scatter in the average SHMR \citep{2010MNRAS.402.1942C,2013MNRAS.433..515W,2014MNRAS.443.3044Z,2015MNRAS.452.1958H, 2016MNRAS.457.3200M, 2021NatAs.tmp..138C}. Similarly, for HOD, the galaxy occupation number may also depend on halo properties such as halo concentration and environment other than the halo mass \citep{2020MNRAS.493.5506H}. The best secondary parameter for the halo characterization is still being sought in the development of HOD and AM .

Stellar mass function (SMF) and galaxy clustering (GC) are the two most commonly used properties to constrain the parameters in HOD and AM. The measurement of these quantities usually relies on spectroscopic surveys with redshift information. In the past two decades, there has been significant progress in large spectroscopic surveys \citep{2000AJ....120.1579Y,2001MNRAS.328.1039C, 2003ApJ...592..728S, 2005A&A...439..845L, 2012ApJS..203...21A, 2012AJ....144..144B, 2014A&A...562A..23G, 2014PASJ...66R...1T, 2016arXiv161100036D, 2020ApJS..249....3A}. In the local universe ($z_s\sim0$), large spectroscopic surveys, in particular the Two Degree
Field Galaxy spectroscopic survey (2dFGRS; \citet{2001MNRAS.328.1039C}) and  the Sloan Digital Sky Survey (SDSS; \citet{2000AJ....120.1579Y}), have been used to measure the SMF and GC down to $10^{9.0}{\rm M_\odot}$ \citep{2001MNRAS.326..255C,2002MNRAS.332..827N,2006MNRAS.368...21L,2009MNRAS.398.2177L,2010ApJ...721..193P}, although the accuracy of the measurements, especially the GC, is still very limited by the survey volume for faint galaxies.  At a higher redshift ($z_s=0.5\sim1.0$), the DEEP2 Galaxy spectroscopic survey \citep{2003SPIE.4834..161D}, the VIMOS-VLT Deep Survey (VVDS; \citet{2005A&A...439..845L}) and the VIMOS Public Extragalactic spectroscopic survey (VIPERS; \citet{2014A&A...562A..23G}) have been used to successfully measure SMF and GC for galaxies with $M_{*}>10^{10.0}{\rm M_\odot}$ \citep{2007A&A...474..443P,2008A&A...478..299M,2013ApJ...767...89M,2013A&A...557A..17M,2013A&A...558A..23D}. However, despite the huge efforts, measurement for fainter objects is still very difficult, and stellar mass limited samples are usually very small at even higher redshift. Fortunately, huge next generation spectroscopic surveys are being constructed for cosmological studies at intermediate to high redshift \citep{2011arXiv1110.3193L,2014PASJ...66R...1T,2016arXiv161100036D}.  Due to limited wavelength coverage and sensitivity of the spectrographs, different populations of galaxies, such as Emission Line Galaxy (ELG), QSOs and Lyman Break Galaxy (LBG), are targeted at different redshifts. These populations are all expected to trace large-scale structures or the Cosmic Webs, so they can be used to extract information for cosmological studies. However, it is difficult to use these surveys to construct stellar mass limited samples for the target selections used by the surveys.

Compared to spectroscopic surveys, photometric surveys, which take the images and obtain the photometric information, are usually deeper and more complete in terms of the stellar mass. However, without precise redshift measurement, the usefulness of photometric surveys is limited. People attempt to infer the photometric redshift from their broad band magnitudes in the photometric surveys \citep{2006A&A...457..841I, 2009ApJ...690.1236I, 2014MNRAS.445.1482S, 2020arXiv200301511N}. Photo-z has been used to measure SMF \citep{2006A&A...459..745F,2008ApJ...675..234P,2010ApJ...709..644I,2012A&A...545A..23B,2021MNRAS.503.4413M} and GC \citep{2016MNRAS.455.4301C,2020ApJ...904..128I, 2021SCPMA..6489811W}, though one has to be very careful about the error and systematics of photo-z especially for faint galaxies. There will be the next generation large and deep multi-band photometric surveys, such as Vera C. Rubin Observatory Legacy Survey of Space and Time (LSST; \citet{2019ApJ...873..111I}) and Euclid \citep{2011arXiv1110.3193L}, which are expected to fairly sample galaxies to an unprecedented faint limit.

As mentioned above, photometric and spectroscopic surveys both have their advantages and disadvantages. Future large spectroscopic surveys have the precise redshift information but only for bright objects and/or selected (or biased) populations, while photometric surveys are deeper and more complete but without accurate redshift measurement. So far, the measurement of SMF and GC is mainly from the spectroscopic surveys. Thus, studies based on SMF and GC, such as HOD and AM, are focused on the local universe or the massive end at higher redshift, resulting in a poor understanding of the faint end and the redshift evolution. To study small and faint objects, quite a few studies attempted to combine a spectroscopic survey with a photometric survey, which can reach several magnitudes fainter than pure spectroscopic surveys. With the Cosmic Webs traced by objects in spectroscopic surveys, properties and distribution of photometric objects around the Cosmic Web can be studied using the photometric surveys. Results can be achieved by stacking satellite and neighbor counts around a large sample of central galaxies, with foreground and background sources subtracted statistically \citep{1987MNRAS.229..621P,1994MNRAS.269..696L,2011ApJ...734...88W,2012MNRAS.427..428G,2013arXiv1303.4722M,2015APh....63...81N,2016MNRAS.459.3998L}. However, most previous studies only use luminosity and color properties of a photometric sample, which is not suitable for quantitative HOD and AM studies, which requires physical properties of galaxies such as stellar mass, rest-frame color and star formation rate (SFR).  

In this paper, we provide a method to measure the distributions and properties (luminosity, mass, color, SFR, morphology et.al.) of Photometric objects Around Cosmic webs (PAC) represented by spectroscopic objects in a spectroscopic survey. The basic idea is that for spectroscopic source $i$ at redshift $z_{s,i}$, only those objects in the photometric sample at around $z_{s,i}$ are correlated the source $i$ and share the similar redshift. Thus for the source $i$, we calculate the physical properties for all sources in the whole photometric sample by assuming they were all at the redshift $z_{s,i}$. Through the cross-correlation of the photometric and spectroscopic samples, foreground or background galaxies with wrong redshift information can be canceled out with the help of random samples, and the true distribution of photometric sources with specified properties around the spectroscopic sources can be obtained. Because both the spectroscopic and photometric samples are huge in the next generation surveys, we will develop a method to speed up the computation for the physical properties.

We introduce the details of PAC in Section \ref{sec:method}. In Section \ref{sec:appli}, we apply PAC to observations. The measurement is modeled in N-body simulation using AM in Section \ref{sec:simu}. And a brief conclusion is given in Section \ref{sec:con}. We adopt the cosmology with $\Omega_m = 0.268$, $\Omega_{\Lambda} = 0.732$ and $H_0 = 71{\rm \ km/s/Mpc}$ through out the paper.

\section{METHODOLOGY}\label{sec:method}
In this section, we introduce a method for estimating $\bar{n}_2w_p(r_p)$ from $w_{12}(\vartheta)$, where $w_p(r_p)$ and $w_{12}(\vartheta)$ are the projected cross-correlation function (PCCF) and the angular cross-correlation function (ACCF) between a given set of spectroscopically identified galaxies and a large sample of photometric galaxies, and $\bar{n}_2$ is the mean number density of the photometric galaxies. The quantity $\bar{n}_2w_p(r_p)$ has clear physical meaning that it measures the {\it true} excess of the photometric objects around the spectroscopic objects on the sky projection. If we choose photometric galaxies at a given stellar mass, the measurement over a range of the stellar mass gives the information on the stellar mass function and the clustering as a function of the stellar mass, which are the key ingredients for understanding the connection of galaxies to dark matter halos.  Therefore, we extend this method to statistically measuring the distribution of the photometric galaxies with specified physical properties (i.e stellar mass, SFR, color etc) around spectroscopically identified galaxies. 

\subsection{Estimating $\bar{n}_2w_p(r_p)$ from $w_{12}(\vartheta)$}
Throughout this section, we call a spectroscopic sample population 1 and a photometric sample population 2.

Assuming an object in population 1 is at distance $r_1$, the number of objects $dN_2$ in population 2 within a solid angle element $d\Omega_2$ in the direction $\bm{r}_2$ is:
\begin{equation}
    dN_2 = \int n_2(\bm{r}_2)r_2^2dr_2d\Omega_2.
\end{equation}
Where $n_2$ is the mean number density of population 2 at the distance $r_2$. The expected number of population 2 objects around a population 1 object is:
\begin{align}
    \langle dN_2\rangle &= \int \langle n_2(\bm{r}_2)\rangle_{1}r_2^2dr_2d\Omega_2 \notag\\
    &= d\Omega_2\int \bar{n}_2[1+\xi_{12}(r_{12})]r_2^2dr_2 \notag\\
    &\approx d\Omega_2[\bar{S}_2+\bar{n}_2w_p(r_1\theta)r_1^2] \,\,.
\end{align}
$\bar{S}_2$ is the mean angular surface density of population 2, $\xi_{12}$ is the cross-correlation function (CCF) between the two populations and $w_p(r_p=r_1\theta)\equiv\int \xi_{12}(\sqrt{r_p^2+\pi^2})d\pi$ is the projected cross-correlation function. The approximation holds if $\theta$ is small. Then we have:
\begin{align}
     \bar{n}_2w_p(r_1\theta)r_1^2 &= \frac{\langle dN_2\rangle}{d\Omega_2}-\bar{S}_2 \notag\\
     &= \bar{S}_2\frac{\langle dN_2\rangle-\langle dR_2\rangle}{\langle dR_2\rangle} \notag\\ 
     &\equiv \bar{S}_2\frac{\langle D_1D_2\rangle-\langle D_1R\rangle}{\langle D_1R\rangle} \notag\\ 
     &= \bar{S}_2w_{12}(\vartheta)\,\,.\label{eq:3}
\end{align}
Here $\langle D_1D_2\rangle$ and $\langle D_1R\rangle$ are the cross pair counts between population 1 and population 2, and between population 1 and the random sample of population 2.

Hence, with our method, $\bar{n}_2w_p(r_1\theta)r_1^2$ can be estimated from $\bar{S}_2w_{12}(\vartheta)$ with only the redshift of population 1. The physical meaning of the quantity $\bar{n}_2w_p(r_1\theta)$ is the excess number of population 2 around population 1.

\subsection{Physical properties of photometric sources around spectroscopically identified sources}
To obtain physical properties for galaxies, distance or redshift information is needed. Unfortunately, wide deep photometric surveys do not have measured redshift for most of galaxies. Photometric redshift $z_p$ is often used to approximate the distance of galaxies, but the errors of $z_p$ are very difficult to estimate especially for faint galaxies. In our method we do not use any information of $z_p$. Instead, since those photometric sources correlated with the spectroscopically identified galaxies must share the same redshift, we can use the spectroscopic redshift for the physical properties for photometric sources around the spectroscopic object. As $\bar{n}_2w_p(r_1\theta)$ is just the number excess  of neighbors around population 1, we can extend our method to estimating the distributions and properties of Photometric objects Around Cosmic webs (PAC) delineated by a spectroscopic survey.

Assuming we only have a spectroscopic sample (population 1) and a large photometric sample, and we want to calculate $\bar{n}_2w_p(r_1\theta)$ between those sources in population 1 at a {\it single} redshift $z_{s}$ and sources with physical property X included in the photometric sample. To select population 2, we assume the whole photometric sample is at redshift $z_{s}$ and calculate their physical properties using SED fitting. As one may note, the calculation is correct only for sources around population 1.  The physical properties calculated are incorrect for foreground and background sources. However, since foreground and background sources along the line-of-sight (LOS) to population 1 are distributed  {\it statistically in the same way} as those along a random LOS, the foreground and background will be canceled out when we calculate $w_{12}(\vartheta)$ (see \citet[Sec 3.3.1]{2011ApJ...734...88W} for a more rigorous derivation). Therefore, population 2 is just selected as the sources with physical property X in the photometric sample even if we assume the whole sample is at redshift $z_s$ in the calculation.
With this method, we can study the distribution of satellites and neighbors with specified physical properties around spectroscopically identified galaxies.

\subsection{Spectroscopic sample with a redshift distribution}

Till now, we have assumed that all the sources in population 1 have the same redshift. However, in reality, spectroscopic samples all have a redshift distribution. If the redshift range is relatively narrow and the evolution of the universe can be neglected, $\bar{n}_2w_p(r_p)$ will not vary much while the change of $r_1$ and $\theta=r_p/r_1$ may not be negligible from one spectroscopic source to another. Thus, $\bar{n}_2w_p(r_p)$ is a better statistic quantity than $\bar{n}_2w_p(r_p)r_1^2$. We re-derive the Equation \ref{eq:3}:
\begin{align}
    \bar{n}_2w_p(r_1\theta) &= \frac{\langle dN_2\rangle/r_1^2}{d\Omega_2}-\frac{\bar{S}_2}{r_1^2} \notag\\
    &= \frac{\langle dN_2\rangle/r_1^2-\langle dR_2\rangle/r_1^2}{\langle dR_2\rangle/r_1^2}\frac{\bar{S}_2}{r_1^2} \notag\\
    &\equiv \frac{\langle D_1D_2\rangle_w-\langle D_1R\rangle_w}{\langle D_1R\rangle_w}\frac{\bar{S}_2}{r_1^2} \notag\\
    &= \frac{\bar{S}_2}{r_1^2}w_{12,weight}(\vartheta)\,\,.
\end{align}
Here $\langle D_1D_2\rangle_w$ and $\langle D_1R\rangle_w$ are the weighted cross pair counts between population 1 and
population 2, and between population 1 and the random sample. Given a set of sources in population 1 with positions $\{\bm{r}_{1,i}\}_{i=1,N_1}$, in order to measure the PCCF at the projected separation $r_p$, we can measure the number $dN_{2,i}$ of population 2 neighbors within the angular separation $\theta \pm 1/2d\theta$ around object $i$ in population 1. Then we measure the weighted count:
\begin{equation}
    \langle D_1D_2\rangle_w = \sum_{i=1,N_1}dN_{2,i}/r_{1,i}^2.
\end{equation}
Similarly we can get the weighted count for random sample:
\begin{equation}
    \langle D_1R\rangle_w = \sum_{i=1,N_1}dR_{2,i}/r_{1,i}^2.
\end{equation}
In an alternative way, we can also change the angular separation used to measure the number counts and make $r_p=r_{1,i}\theta_{i}$ the same for each galaxy in population 1. Then, we sum the counts around each galaxies without weighting. These two methods are equivalent. 

To minimize the effects of complex survey geometries, we adopt an estimator in analogy to the Landy-Szalay estimator for two-point auto-correlation \citep{1993ApJ...412...64L}:
\begin{equation}
  w_{12,weight}(\theta) = \frac{\langle D_1D_2\rangle_w-\langle D_1R_2\rangle_w-\langle D_2R_1\rangle_w+\langle R_1R_2\rangle_w}{\langle R_1R_2\rangle_w},
\end{equation}
where $R_1$ and $R_2$ are the random points for spectroscopic and photometric samples respectively.

The above method has accounted for the variance of $\theta$ with redshift for sources in population 1, while $r_1$ in $\bar{S}_2w_{12,weight}/r_1^2$ still varies with redshift. Therefore, we divide population 1 into narrower redshift bins and reduced the error from the change of $r_1$. 

For population 1 with a redshift distribution, we can divided them into $m$ redshift bins. $m$ can be as large as possible only if there are sufficient galaxies in each bin. If the mean redshifts for these bins are $\{z_{s,j}\}_{j=1,m}$, we calculate the physical properties for the whole photometric catalog and select a population $2$ for each $z_{s,j}$. Then, we calculate $\bar{S}_{2,j}w_{12,weight,j}(\vartheta)/r_{1,j}^2$ for each redshift bin, and the mean $\bar{n}_2w_p(r_1\theta)$ of population 1 can be obtained by averaging over these redshift bins:
\begin{equation}
    \bar{n}_2w_p(r_1\theta) = \frac{1}{m}\sum_{j=1}^{m}\frac{\bar{S}_{2,j}}{r_{1,j}^2}w_{12,weight,j}(\vartheta).\label{equ:nw}
\end{equation}

\begin{figure*}
\plotone{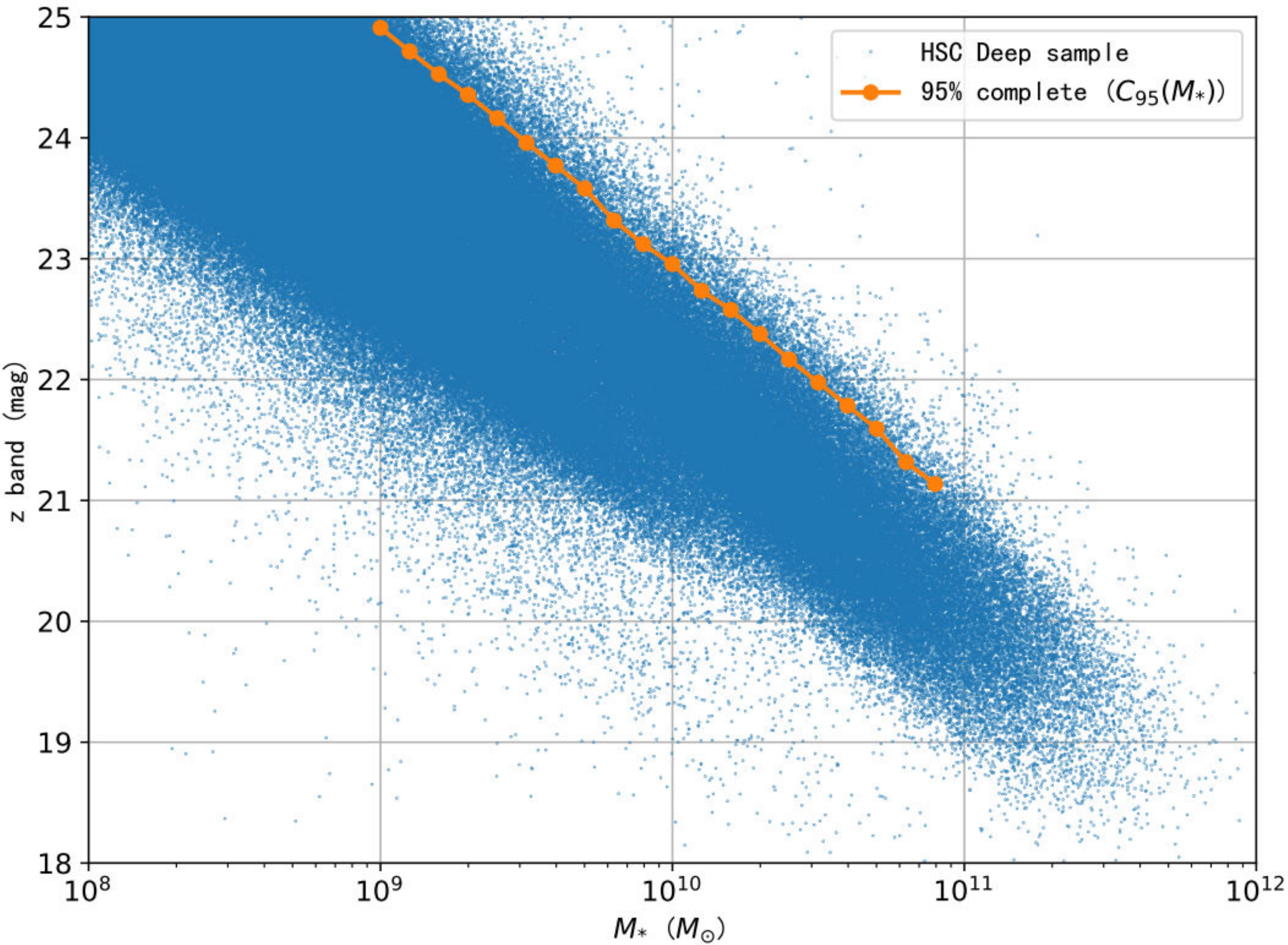}
\caption{The stellar mass-z-band magnitude relation for HSC deep field galaxies with photo-z between 0.5 and 0.7. The blue dots are galaxies in the sample and the orange line shows the $95\%$ completeness limit $C_{95}(M_{*})$ of the z-band magnitude. \label{fig:f1}}
\end{figure*}

\section{APPLICATIONS TO CMASS and HSC samples}\label{sec:appli}
In this section, we apply PAC to CMASS spectroscopic sample in the Baryon Oscillation Spectroscopic Survey (BOSS; \citet{2012ApJS..203...21A}; \citet{2012AJ....144..144B}) and Hyper
Suprime-Cam Subaru Strategic Program (HSC-SSP; \citet{2019PASJ...71..114A}) PDR2 wide field photometric sample.

\begin{figure*}
\plotone{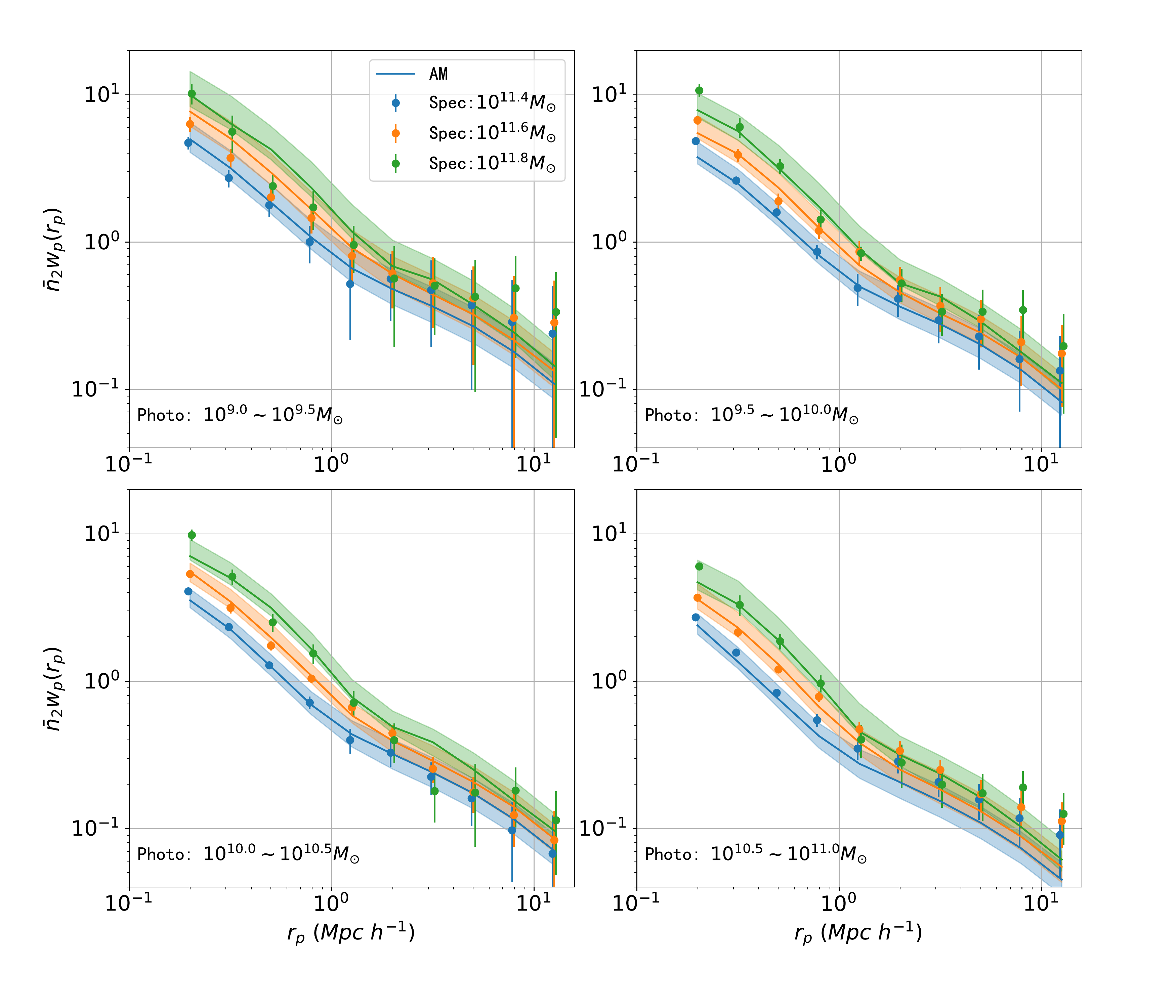}
\caption{Our measurement of $\bar{n}_2w_p(r_p)$ for eleven mass bins. Each panel shows the results for one mass bin of photometric sample. Colored dots with error bars are the results from observation. Lines with shadows are the median results and $1\sigma$ error of our modeling from the abundance matching (AM) using MCMC sampling method. \label{fig:f2}}
\end{figure*}

\begin{figure}
    \plotone{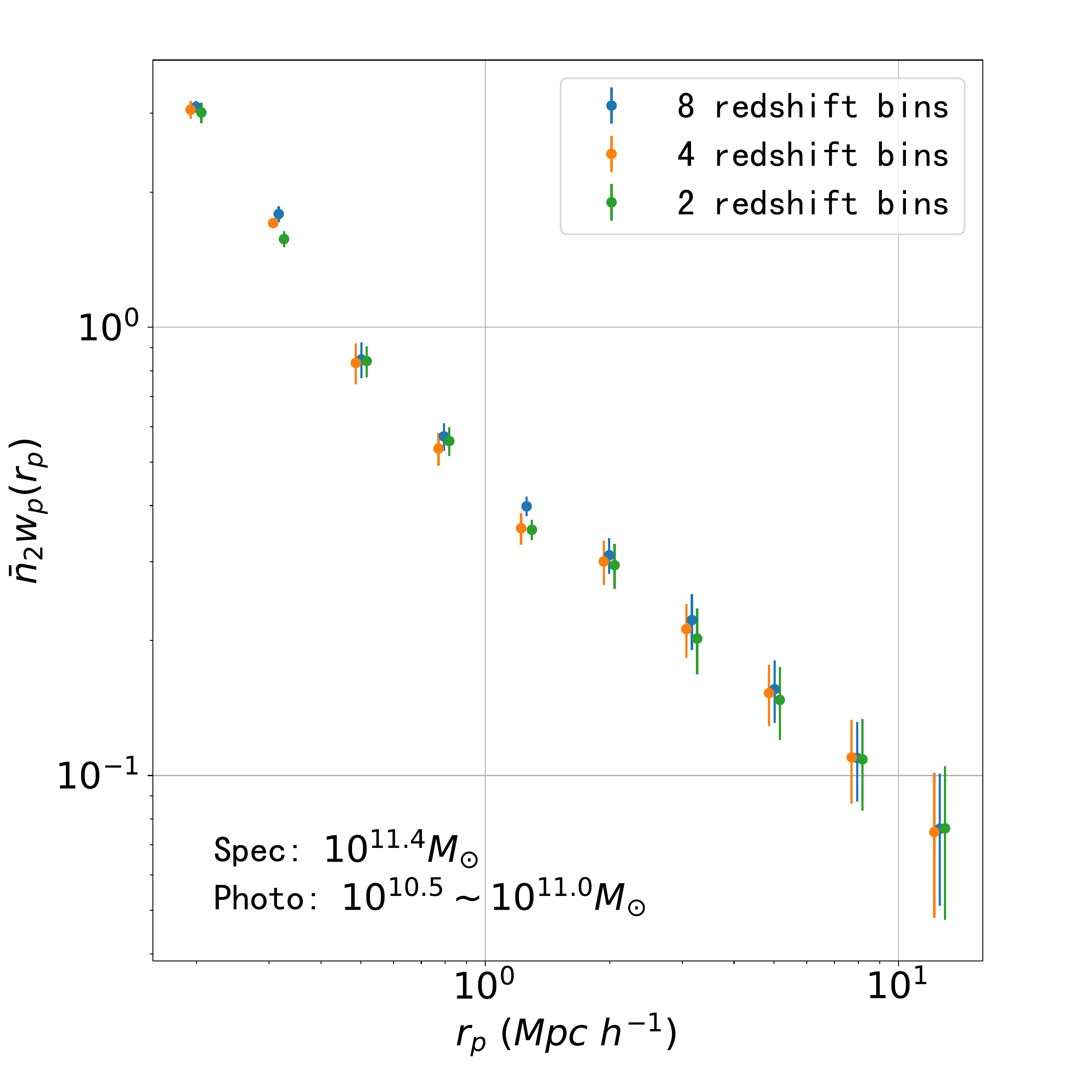}
    \caption{Comparison of  the  $\bar{n}_2w_p(r_p)$ measured with different divisions of redshift bins (as indicated in the figure) for the spectroscopic sample. The result is insensitive to the number of redshift bins used}
    \label{fig:bins}
\end{figure}

\subsection{Observational data}
We use the HSC-SSP PDR2 wide field photometric catalog \citep{2019PASJ...71..114A} as the photometric sample. To obtain more accurate physical properties, we choose sources in the footprints observed with all five bands ($grizy$) to ensure that there are enough bands for SED. Sources around bright objects are masked using {\texttt{\{grizy\}\_mask\_pdr2\_bright\_objectcenter}} flag provided by the HSC collaboration \citep{2018PASJ...70S...7C}. And we use the {\texttt{\{grizy\}\_extendedness\_value}} flag to exclude stars in the sample. Finally, there are around $2\times10^8$ galaxies in our photometric sample. We can also construct a random point catalog ($100/arcmin^2$) with the same selection criteria from the HSC database for ACCF analysis. The effective area calculated from the random point number is $501\ deg^2$.

To ensure the distribution of foreground and background galaxies is the same for all LOS directions, the easiest way is to make sure that the sample is complete for galaxies with specified physical properties at the required redshift. Since the survey depth is not uniform across the HSC survey, for a low stellar mass limit, some patches remain complete while others may not. Therefore, we use HSC-SSP PDR2 deep field catalog and {\texttt{DEmP}} photo-z \citep{2020arXiv200301511N} to study the completeness of galaxies for different stellar mass. We select galaxies with photo-z between $0.5$ and $0.7$ and calculate the physical properties for these galaxies with five bands $grizy$ using the SED code {\texttt{CIGALE}} \citep{2019A&A...622A.103B}. The stellar population synthesis models of
\citet{2003MNRAS.344.1000B} are used to compute the physical properties of
galaxies. In these calculations, the \citet{2003PASP..115..763C} Initial Mass
Function is adopted. We assume a delayed star formation history $\phi(t)\approx t\exp{-t/\tau}$ where $\tau$ is taken from $10^7$ to $1.258\times
10^{10}$ yr with an equal logarithmic interval $\Delta lg \tau=0.1$. Three metallicities, $Z/Z_{\odot}=0.4$, 1, and 2.5, are considered, where $Z_{\odot}$ is the metallicity of the Sun. We use the extinction law of \citet{2000ApJ...533..682C} with dust reddening in the range $0<E(B-V)<0.5$. As shown in Figure \ref{fig:f1}, stellar mass shows a clear correlation with $z$ band magnitude at $0.5<z_p<0.7$, so magnitude limit can be used to derive a complete sample for specified stellar mass. Particularly, we define the z-band completeness limit $C_{95}(M_{*})$ that $95\%$ of the galaxies are brighter than $C_{95}(M_{*})$ in the z-band for given stellar mass $M_{*}$. (orange line in Figure \ref{fig:f1}). Therefore, for stellar mass bin $[m_l,m_h]$ of interest, we only use the data in the survey footprints deeper than $C_{95}(m_l)$, where the depth is defined as the $z$ band limiting magnitude of $10\sigma$ detection  for a point source. We would like to stress again that the photo-z is used only here to find the completeness limit, and it is not used in the following clustering analysis. 

We use the CMASS sample in BOSS that consists of massive galaxies with $i<19.9\ mag$ as spectroscopic sample (population 1). We first select galaxies in the redshift range of $0.5<z_s<0.7$, and then cross match it with the HSC photometric sample we constructed above and The DESI Legacy Imaging Surveys DR9 catalog \citep{2019AJ....157..168D}. After that, we get the magnitudes in seven bands $grizyW1W2$ for each CMASS galaxy in the footprint of HSC. We calculate the physical properties for these galaxies using the same SED templates but with seven bands $grizyW1W2$. As noticed by previous
studies \citep{2013MNRAS.435.2764M,2016MNRAS.457.4021L,2018ApJ...858...30G}, the CMASS sample is complete
to stellar mass $M_{*}\approx 10^{11.3}{\rm M_\odot}$. Therefore, we adopt a stellar mass cut at $10^{11.3} {\rm M_\odot}$ in this study. Moreover, we only consider central galaxies in the spectroscopic sample, so we select galaxies which do not have more massive neighbors within the projected distance of $1Mpc\ h^{-1}$. Finally, there are $8028$ massive ($>10^{11.3}{\rm M_\odot}$) central galaxies left in the spectroscopic sample.

\begin{figure}
    \plotone{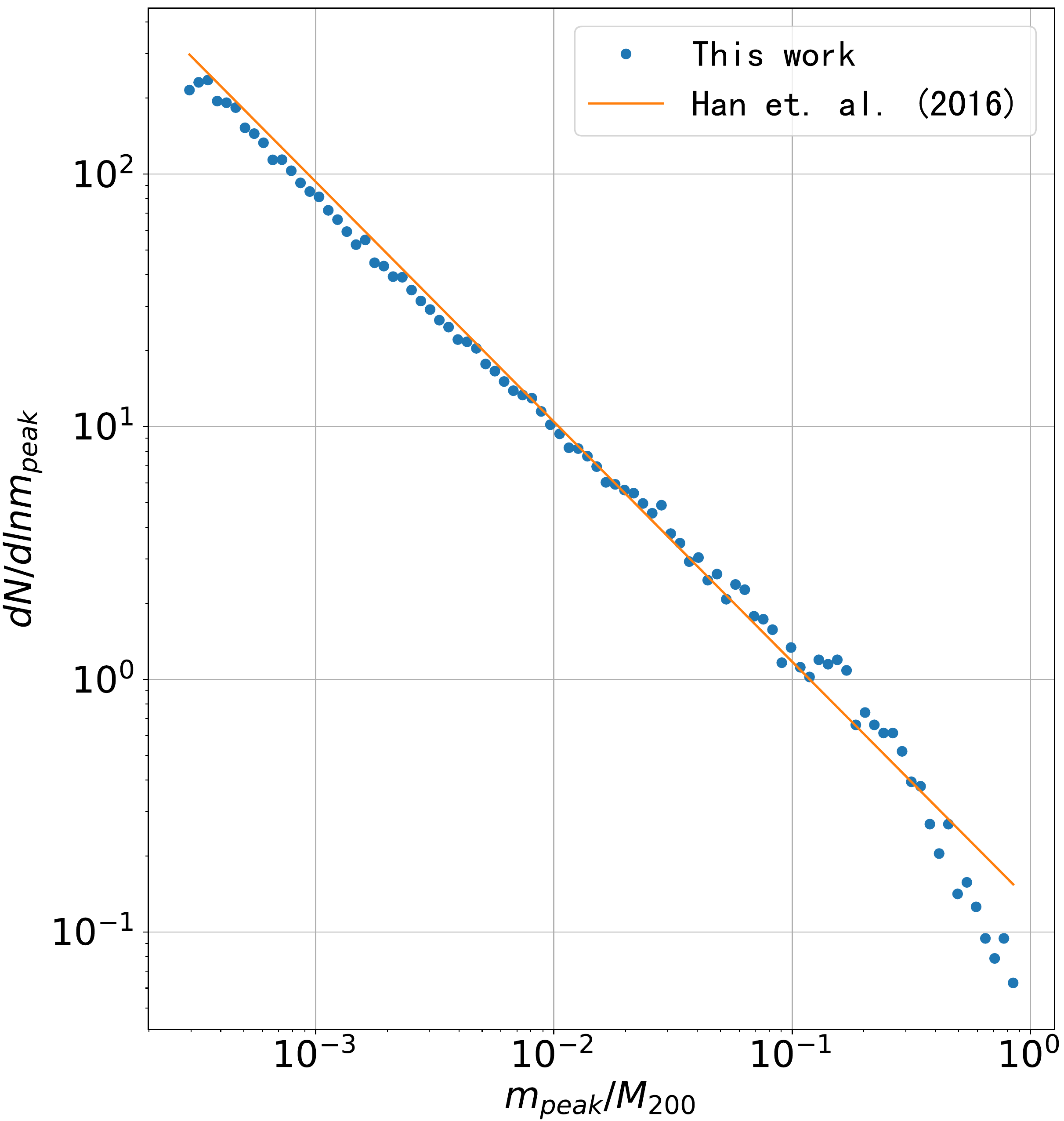}
    \caption{Subhalo mass ($m_{peak}$) function in halos with $M_{200}=10^{14.0}{\rm M_\odot} h^{-1}$. Dots show the results from CosmicGrowth Simulation after correction for subhalos with less than $20$ particles. Solid line shows the result from \citet{2016MNRAS.457.1208H} based on high resolution zoom-in simulations.}
    \label{fig:f3}
\end{figure}

\begin{figure*}
\plotone{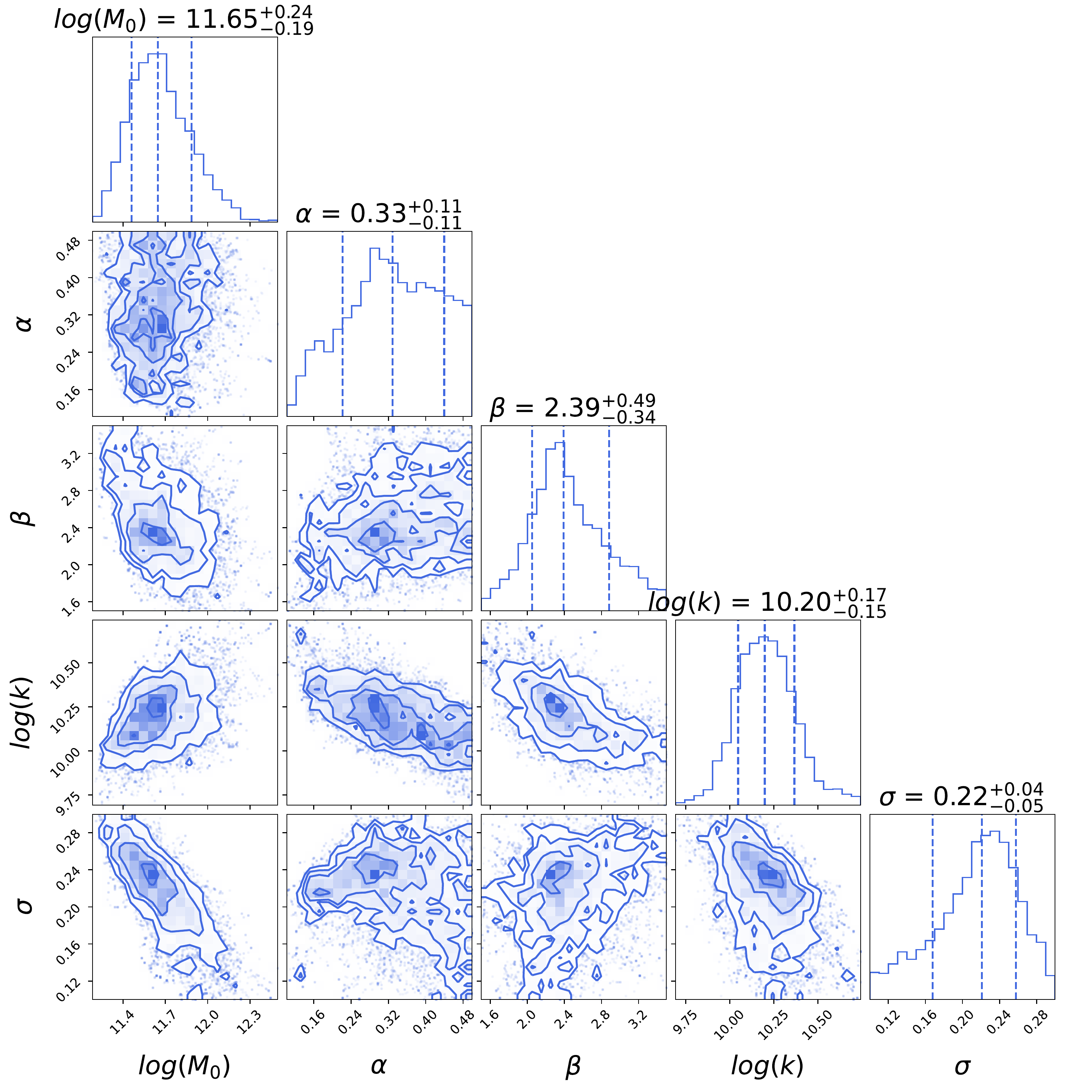}
\caption{Constraints on parameters of the SHMR model using MCMC sampling. The central value is a median and the error means $16\sim84$ percentiles after other parameters are marginalized over. \label{fig:f4}}
\end{figure*}

\begin{deluxetable*}{cccccc}
\label{tab:t1}
\tablenum{1}
\tablecaption{Posterior PDFs of the parameters from MCMC for the SHMR model.}
\tablewidth{0pt}
\tablehead{
\colhead{model}&\colhead{$M_0$}&\colhead{$\alpha$}&\colhead{$\beta$}&\colhead{$k$}&\colhead{$\sigma$}\\
\colhead{}&\colhead{(${\rm M_\odot}\ h^{-1}$)}&\colhead{}&\colhead{}&\colhead{(${\rm M_\odot}$)}&\colhead{}
}
\startdata
Unified&$10^{11.65^{+0.24}_{-0.19}}$&$0.33^{+0.11}_{-0.11}$&$2.39^{+0.49}_{-0.34}$&$10^{10.20^{+0.17}_{-0.15}}$&$0.22^{+0.04}_{-0.05}$\\
\enddata
\end{deluxetable*}

\subsection{PAC for different mass bins}
We first divide the spectroscopic sample into three mass bins $[10^{11.3},10^{11.5},10^{11.7},10^{11.9}]\ {\rm M_\odot}$. In each mass bin, galaxies are divided into four redshift bins $[0.5,0.55,0.6,0.65,0.7]$. Then, we perform SED for the whole photometric sample at redshifts 0.525, 0.575, 0.625 and 0.675 respectively. After that, the photometric sample is also divided into four mass bins $[10^{9.0},10^{9.5},10^{10.0},10^{10.5},10^{11.0}]\ {\rm M_\odot}$ at each redshift. Using Equation \ref{equ:nw}, $\bar{n}_2w_p(r_p)$ for twelve mass bins (three spectroscopic $\times$ four photometric) in the redshift range $0.5<z_s<0.7$ can be obtained. During the calculation, only footprints deeper than $C_{95}(m_l)$ are used for each mass bin. 

\begin{figure*}
\plottwo{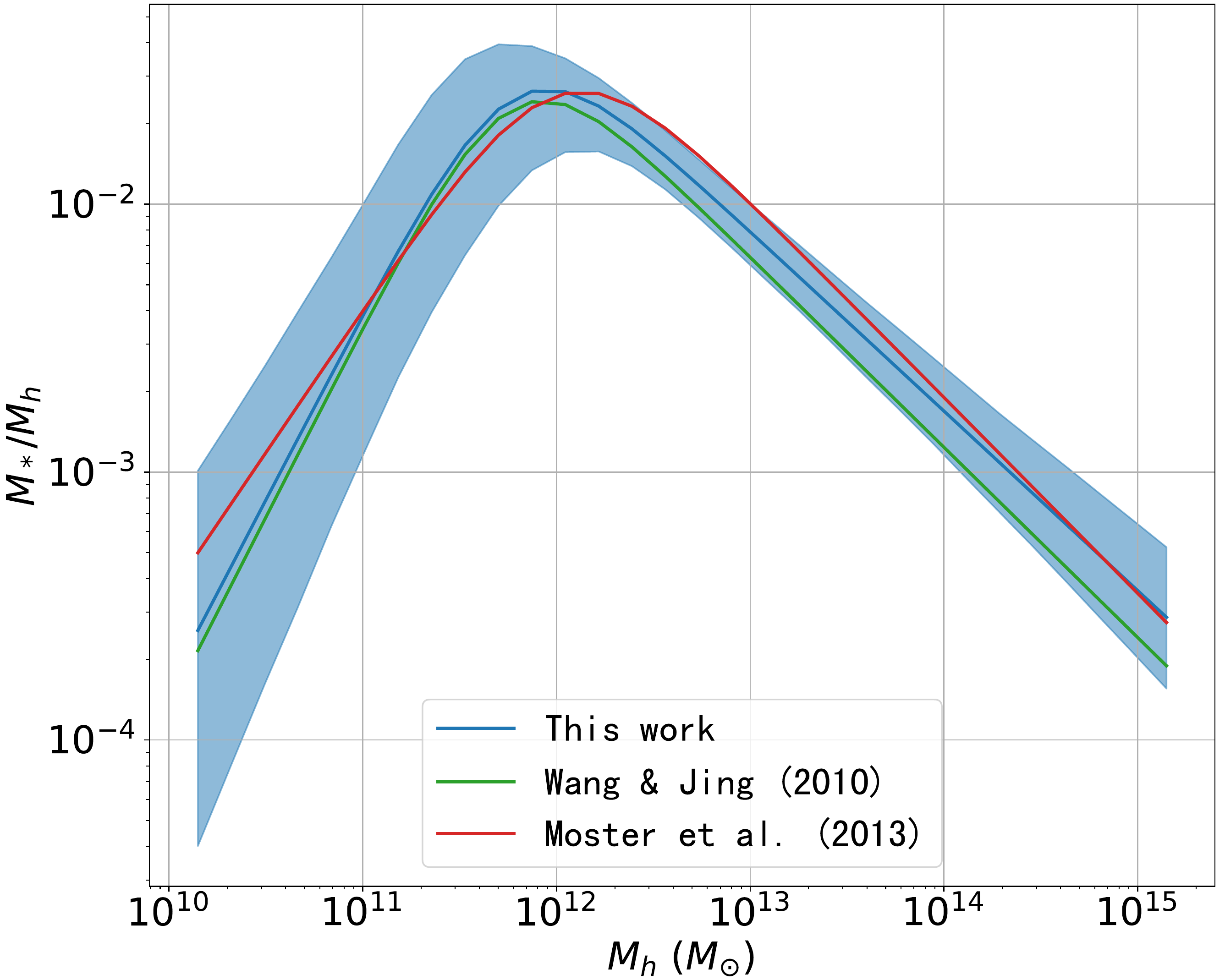}{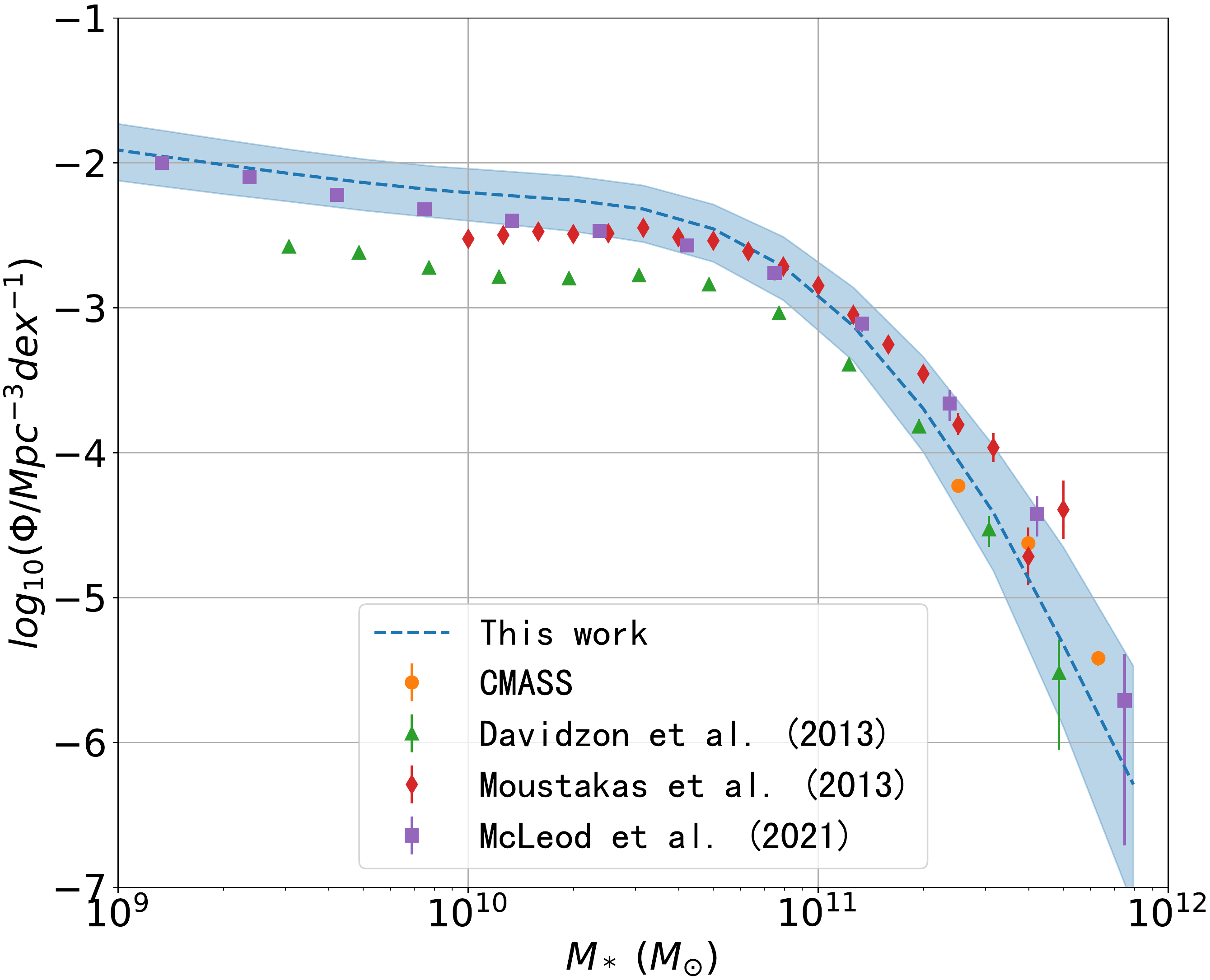}
\caption{Left: The mean SHMR and its error from our work ($0.5<z_s<0.7$) comparing with the results from \citet{2010MNRAS.402.1796W} ($z_s\sim0.8$) and \citet{2013MNRAS.428.3121M} ($z_s=0.6$). Blue line with shallow shows the SHMR from our work. Green and red lines show the results from \citet{2010MNRAS.402.1796W} and \citet{2013MNRAS.428.3121M}. Right: SMF from AM comparing to observations. Blue dashed line with shadow shows the results from our work. Orange dots show the measurement at high mass end from our CMASS spectroscopic sample. Green triangles and red diamonds show the measurements from VIPERS ($0.50<z_s<0.60$; \citet{2013A&A...558A..23D}) and PRIMUS ($0.50<z_s<0.65$; \citet{2013ApJ...767...50M}) with spectroscopic redshift, and purple squares show the results from \citet{2021MNRAS.503.4413M} ($0.25<z_p<0.75$) measured in various deep photometric surveys with photo-z ($z_p$).\label{fig:f5}}
\end{figure*}

To evaluate the statistical error, we use the Jackknife resampling technique \citep{1982jbor.book.....E}. Mean value and error of the mean value of $\bar{n}_2w_p(r_p)$ for each mass bin can be calculated as:
\begin{equation}
    \bar{n}_2w_p(r_p) = \frac{1}{N_{\rm sub}}\sum_{k=1}^{N_{\rm sub}}\bar{n}_{2,k}w_{p,k}(r_p)
\end{equation}
\begin{equation}
    \sigma^2 = \frac{N_{\rm sub}-1}{N_{\rm sub}}\sum_{k=1}^{N_{\rm sub}}(\bar{n}_{2,k}w_{p,k}(r_p)-\bar{n}_2w_p(r_p))^2,
\end{equation}
where $N_{\rm sub}$ is the number of jackknifed realizations and $\bar{n}_{2,k}w_{p,k}(r_p)$ is the excess of projected density of the $k$th realization. In this work, we adopt $N_{\rm sub}=50$.

The results are shown in Figure \ref{fig:f2} with colored dots. Each panel shows the results for the same mass bin of the photometric sample. Since $\bar{n}_2$ is the same in the same panel, the difference of $\bar{n}_2w_p(r_p)$ reflects the difference of $w_p(r_p)$ between different spectroscopic mass bins. Though the footprint become smaller for the lowest mass bin  $10^{9.0}{\rm M_\odot}$, the clustering signal is still quite good for all the three spectroscopic subsamples, so we can study the properties such as SMF and SHMR down to the low mass end.

To test the robustness of PAC to the redshift bin used, we compare the results when different number of redshift bins are used to divide the spectroscopic subsample. As we can see from Figure \ref{fig:bins}, for a narrow redshift range as in our study ($0.5<z<0.7$), the measurement is nearly independent of the number of redshift bins used, indicating that our algorithm is robust. 

\section{Abundance matching with N-body simulation}\label{sec:simu}
With the density and clustering information of galaxies, we can study the galaxy–halo connection using N-body simulation. HOD and AM are the two most commonly used methods for populating galaxies to dark matter halos. We follow \citet{2010MNRAS.402.1796W} and use AM to study the galaxy–halo relation based on $\bar{n}_{2}w_{p}(r_p)$ for different mass bins obtained in the observation. As we will show, the stellar-halo mass relation (SHMR), the stellar mass function (SMF) and the conditional stellar mass function (CSMF) for satellites can be inferred from the AM results for galaxies in a wide range of stellar mass ($10^{9.0}-10^{12.0}{\rm M_\odot}$).

\subsection{CosmicGrowth Simulation}
We use the CosmicGrowth Simulation \citep{2019SCPMA..6219511J} for our studies. CosmicGrowth Simulation is a grid of high resolution N-body simulations run in different cosmologies using an adaptive parallel P$^3$M code \citep{2002ApJ...574..538J}. We use the $\Lambda$CDM simulation with cosmological parameters $\Omega_m = 0.268$, $\Omega_{\Lambda} = 0.732$ and $\sigma_8 = 0.831$. The box size is $600\ Mpc\ h^{-1}$ with $3072^3$ dark matter particles and softening length $\eta = 0.01\ Mpc\ h^{-1}$. Groups are identified with the Friends-of-Friends algorithm with a linking length $0.2$ times the mean particle separation. The halos are then processed with HBT+ \citep{2012MNRAS.427.2437H, 2018MNRAS.474..604H} to obtain subhalos and their evolution histories. We use the catalog of Snapshot 83 at redshift about 0.57 to compare with the observation. 

We also use the fitting formula in \citet{2008ApJ...675.1095J} to evaluate the merger timescale for subhalos with less than $20$ particles (including orphans), which may be unresolved, and abandon those that have already merged into central subhalos. In Figure \ref{fig:f3}, we compared our subhalo mass function in halos with $M_{200}=10^{14.0}{\rm M_\odot} h^{-1}$ to the universal subhalo mass function of \citet{2016MNRAS.457.1208H} who used high resolution zoom-in simulations (mass resolution $\sim10^{3}{\rm M_\odot} h^{-1}$ for the highest one) and carefully corrected for the resolution effect. In this comparison, the halos are defined as objects of the radius $R_{200}$ within which the average density equals 200 times the critical density of the universe, and the subhalo mass is defined as its peak halo mass $m_{peak}$ in its history, to be consistent with \citet{2016MNRAS.457.1208H}. Our subhalo mass function is in good agreement with \citet{2016MNRAS.457.1208H} down to $10^{10.5}{\rm M_\odot} h^{-1}$, which is good enough for this study.

\subsection{Abundance matching}
The SHMR can be described by a formula of double power-law form:
\begin{equation}
    M_{*} = \left[\frac{2}{(\frac{M_{acc}}{M_0})^{-\alpha}+(\frac{M_{acc}}{M_0})^{-\beta}}\right]k\,,
\end{equation}
where $M_{acc}$ is defined as the viral mass $M_{vir}$ of the halo at the time when the galaxy was last the central dominant object. We use the fitting formula in \citet{1998ApJ...495...80B} to find $M_{vir}$. The scatter in $\log(M_*)$ at a given $M_{acc}$ is described with a Gaussian function of the width $\sigma$. We use the same set of parameters for centrals and satellites (unified model) as in many studies \citep{2010MNRAS.402.1796W, 2019MNRAS.488.3143B}. 

Once the parameters $\{M_0,\alpha,\beta,k,\sigma\}$ are fixed, galaxies can be assigned to each dark matter halo. To compare $\bar{n}_2w_p(r_p)$ with observation, we define $\chi^2$ as:
\begin{equation}
    \chi^2 = \frac{1}{N_p}\sum_{N_p}\left[\frac{\log(\bar{n}_2w_p(r_p))_{sim}-\log(\bar{n}_2w_p(r_p))_{ob}}{\sigma(\log(\bar{n}_2w_p(r_p))_{ob})}\right]^2\,,
\end{equation}
where $N_p$ is the total number of points over which $\bar{n}_2w_p(r_p)$ is compared. We only consider the radius range of $0.1<r_p<10\ Mpc\ h^{-1}$, in order to avoid the deblending problem of the HSC catalog at the smaller $r_p$ \citep{2021ApJ...919...25W}, and large errors at $r_p>10Mpc\ h^{-1}$. 
In order to perform a maximum likelihood analysis, we use the Markov chain Monte Carlo (MCMC) sampler {\texttt{emcee}} \citep{2013PASP..125..306F}.

Posterior PDFs of the parameters of the SHMR model from MCMC are shown in Figure \ref{fig:f4} and Table \ref{tab:t1}. And the corresponding $\bar{n}_2w_p(r_p)$ and errors for each mass bin is shown by solid lines with shadows in Figure \ref{fig:f2}. The fitting is overall good for all mass bins in our samples and all the parameters are constrained well. 

\begin{figure}
    \plotone{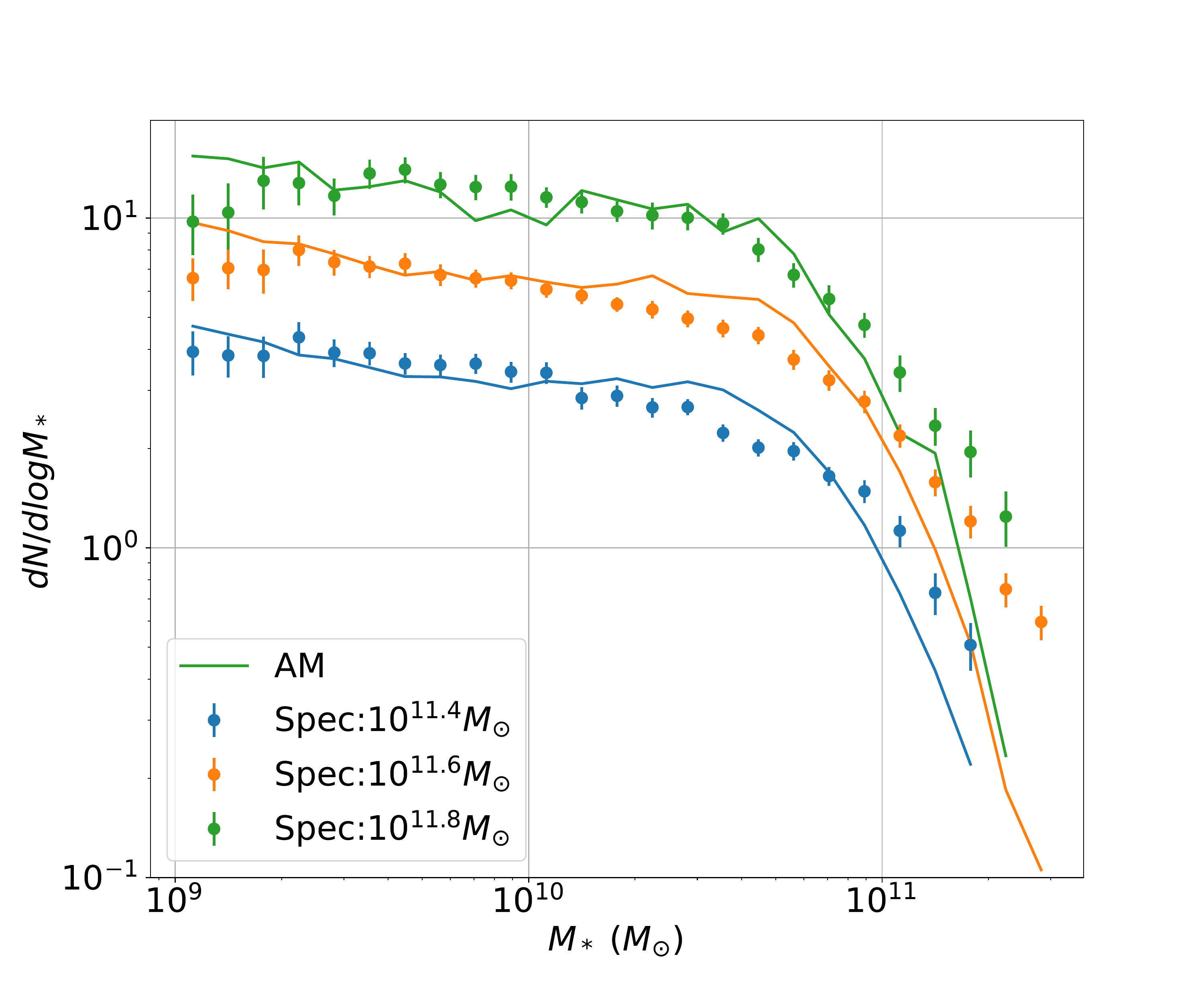}
    \caption{CSMF of satellites for centrals with different mass.}
    \label{fig:csmf}
\end{figure}

\subsection{SHMR and SMF}

After we derive the parameters, we can calculate the SHMR and SMF at redshift $0.5\sim0.7$ for stellar mass range $10^{9.0}\sim10^{12.0}{\rm M_\odot}$ covered by our observational samples. 

In the left panel of Figure \ref{fig:f5}, we compare the SHMR from our results with other works. Blue line with shadow shows the SHMR from our work. Green line shows the unified SHMR model from \citet{2010MNRAS.402.1796W} by fitting the SMF and GC at $z_s\sim0.8$ from VVDS \citep{2007A&A...474..443P,2008A&A...478..299M}. \citet{2013MNRAS.428.3121M} uses a redshift-dependent parametrization of SHMR and constrains the parameters by fitting the SMFs from SDSS \citep{2009MNRAS.398.2177L}, Spitzer \citep{2008ApJ...675..234P} and Wide Field Camera 3 (WFC3) \citep{2012A&A...538A..33S} varying from $z_s\sim4$ to the present. We show their results at $z_s=0.6$ with red line in the figure. The SHMR from our and other works are in good agreement with each other.

SMF from our AM model and other observational measurements are shown in the right panel of Figure \ref{fig:f5}. Blue dashed line with shadow shows the results from AM of our work. Orange dots show our own measurement from the CMASS spectroscopic sample at the high mass end. Green triangles and red diamonds show the measurements from VIPERS ($0.50<z_s<0.60$; \citet{2013A&A...558A..23D}) and the PRism MUlti-object Survey (PRIMUS; $0.50<z_s<0.65$; \citet{2013ApJ...767...50M}) with spectroscopic redshift. And purple squares are the results from \citet{2021MNRAS.503.4413M} ($0.25<z_p<0.75$), in which they measure the SMF using various deep photometric surveys (UKIDSS Ultra Deep Survey (UDS), COSMOS and CFHTLS-D1) with photo-z. At the high mass end, the results from AM and various observational measurements are consistent. While at the low mass end, the discrepancy between observations is still significant (nearly by a factor of two between \citet{2013A&A...558A..23D} and \citet{2013ApJ...767...50M}), which has been exhaustively discussed in literature \citep{2006MNRAS.368...21L,2007A&A...474..443P,2013A&A...558A..23D}. Since it is hard to detect the faint objects and the survey areas of the high redshift deep surveys are usually small, completeness, selection effects and cosmic variance should be carefully considered. Different weighting methods to compensate for the incompleteness and regions with different densities that the surveys observed can produce huge difference in SMF. Therefore, the SMF of the faint end at higher redshift is still very hard to measure, for which PAC is a very promising method that combines the advantages of both the photometric and spectroscopic surveys. Interestingly, the SMF from our work using PAC and AM is well consistent with a very recent measurement of \citet{2021MNRAS.503.4413M} down to $10^{9.0}{\rm M_\odot}$.  

\subsection{CSMF of satellites}
The CSMF of satellites around a central galaxy of given stellar mass can be derived from the PAC measurement and the AM modeling. We define satellite to be the galaxies within $R_{vir}$ of the dark matter halos of the central galaxies. To get the number of satellites, the most straight forward way is to sum up the excess surface density $\bar{n}_2w_p(r_p)$ weighted by area within $R_{vir}$. However, two effects should be corrected in our study. The surface density includes all excess of galaxies along the line-of-sight direction rather than within $R_{vir}$ (see \citet{2012ApJ...760...16J}) and the measurement within $0.1R_{vir}$ is unreliable for the HSC PDR2 photometric catalog due to the deblending problem \citep{2021ApJ...919...25W}.

Therefore, we use the results from AM to compensate these two effects. After populating the galaxies to halos, we measure the average number of satellites within the projected radius range $0.1R_{vir}<r_p<R_{vir}$ and within virial radius $R_{vir}$ for each central and satellite mass bins. Then, we calculate the average number of satellites within $0.1R_{vir}<r_p<R_{vir}$ in observation and infer the number within $R_{vir}$ using the ratio calculated from simulation. We show the CSMF from $10^{9.0}{\rm M_\odot}$ to $10^{11.6}{\rm M_\odot}$ for central galaxies with different mass bins in Figure \ref{fig:csmf}. Colored dots show the results from the observation and the solid lines are the results from AM. The results are consistence with each other at all mass bins for observation and simulation. 

\section{Conclusion}\label{sec:con}
In this paper, we provide a method for estimating the projected density distribution $\bar{n}_2w_p(r_p)$ from  $w_{12}(\vartheta)$ and extending this method to measure the distributions and properties of Photometric objects Around Cosmic web (PAC) traced by spectroscopic surveys. Basically, by assuming the whole photometric sample at the same redshift as spectroscopic sources, we can calculate the physical properties of the photometric sample. And through cross-correlation, foreground and background galaxies with wrong properties are canceled out and the true distribution of photometric sources with specified physical properties around the spectroscopic sources can be obtained. 

We apply PAC to massive ($>10^{11.3}{\rm M_\odot}$) central galaxies in BOSS CMASS sample ($0.5<z<0.7$) and HSC-SSP PDR2 wide field photometric sample. We calculate $\bar{n}_2w_p(r_p)$ for several stellar mass bins (three for CMASS $\times$ four for HSC) from $10^{9.0}{\rm M_\odot}$ to $10^{12.0}{\rm M_\odot}$ and the measurement is overall good at $0.1\ Mpc\ h^{-1}<r_p<10\ Mpc\ h^{-1}$ for all the mass bins. Then, we use abundance matching to model $\bar{n}_2w_p(r_p)$ in N-body simulation with MCMC sampling. We use the same set of parameters for central and satellite galaxies to model the observation. All the parameters are constrained well and the fitting to observation is overall good for all mass bins. Our AM model can accurately reproduce the SMF comparing to observations. The SHMR from our results are also in good agreement with previous works. Using PAC and AM, we also calculate the CSMF of satellites for centrals with different mass. 

We expect that PAC will have many applications with ongoing and upcoming photometric and spectroscopic surveys. However, since the wide spectroscopic surveys at higher redshifts only target specific populations of galaxies like ELGs and QSOs, the galaxy-halo connection of these populations, which is also one of the key challenges for galaxy formation and cosmological studies, should be well established to make full use of PAC. Recently, there has been some progress made in ELG HOD and AM modeling with spectroscopic or narrow band data \citep[H, Gao et al. 2021, in preparation]{2019ApJ...871..147G,2021PASJ...73.1186O}. PAC can provide more information for studying the galaxy-halo connection of ELGs. As in this work, the SHMR of normal galaxies can be obtained using PAC with a stellar mass limited spectroscopic sample such as Large Red Galaxies (LRGs). If the redshift range of ELGs is overlapped with that of LRGs, by applying PAC to ELGs and the same photometric sample, we can obtain galaxy bias of ELGs with respect to the underlying matter distribution. With the better understanding of the galaxy-halo connection for ELGs, we can extend PAC to a higher redshift where LRGs cannot be reached. It is even worth trying to simultaneously study the connections of the ELGs and the normal galaxy population to dark matter halos, since the projected density distribution $\bar{n}_2w_p(r_p)$ is expected to be precisely measured with next generation galaxy surveys. 

With PAC and future surveys, we can study the galaxy-halo connection for galaxies with different physical properties other than mass, such as SFR, color and morphology \citep{2021arXiv211005760X}. We can also push the understanding of SHMR, SMF and other properties to higher redshift and fainter luminosity end. With the properties and distribution of satellite, we can also study the galaxy evolution such as galaxy merger rate, merger timescale and environment quenching. Moreover, since PAC has very strong signal at the small scale, we can also quantify the fiber collision effect in spectroscopic samples. We may also apply the method to photo-z samples to quantify photo-z errors. We will explore these applications in our future studies.

\begin{acknowledgments}

The work is supported by NSFC (12133006, 11890691, 11621303) and by 111 project No. B20019. This work made use of the Gravity Supercomputer at the Department of Astronomy, Shanghai Jiao Tong University.

The Hyper Suprime-Cam (HSC) collaboration includes the astronomical communities of Japan and Taiwan, and Princeton University. The HSC instrumentation and software were developed by the National Astronomical Observatory of Japan (NAOJ), the Kavli Institute for the Physics and Mathematics of the Universe (Kavli IPMU), the University of Tokyo, the High Energy Accelerator Research Organization (KEK), the Academia Sinica Institute for Astronomy and Astrophysics in Taiwan (ASIAA), and Princeton University. Funding was contributed by the FIRST program from Japanese Cabinet Office, the Ministry of Education, Culture, Sports, Science and Technology (MEXT), the Japan Society for the Promotion of Science (JSPS), Japan Science and Technology Agency (JST), the Toray Science Foundation, NAOJ, Kavli IPMU, KEK, ASIAA, and Princeton University. 

This publication has made use of data products from the Sloan Digital Sky Survey (SDSS). Funding for SDSS and SDSS-II has been provided by the Alfred P. Sloan Foundation, the Participating Institutions, the National Science Foundation, the U.S. Department of Energy, the National Aeronautics and Space Administration, the Japanese Monbukagakusho, the Max Planck Society, and the Higher Education Funding Council for England.

The Legacy Surveys consist of three individual and complementary projects: the Dark Energy Camera Legacy Survey (DECaLS; Proposal ID \#2014B-0404; PIs: David Schlegel and Arjun Dey), the Beijing-Arizona Sky Survey (BASS; NOAO Prop. ID \#2015A-0801; PIs: Zhou Xu and Xiaohui Fan), and the Mayall z-band Legacy Survey (MzLS; Prop. ID \#2016A-0453; PI: Arjun Dey). DECaLS, BASS and MzLS together include data obtained, respectively, at the Blanco telescope, Cerro Tololo Inter-American Observatory, NSF’s NOIRLab; the Bok telescope, Steward Observatory, University of Arizona; and the Mayall telescope, Kitt Peak National Observatory, NOIRLab. The Legacy Surveys project is honored to be permitted to conduct astronomical research on Iolkam Du’ag (Kitt Peak), a mountain with particular significance to the Tohono O’odham Nation.

NOIRLab is operated by the Association of Universities for Research in Astronomy (AURA) under a cooperative agreement with the National Science Foundation.

This project used data obtained with the Dark Energy Camera (DECam), which was constructed by the Dark Energy Survey (DES) collaboration. Funding for the DES Projects has been provided by the U.S. Department of Energy, the U.S. National Science Foundation, the Ministry of Science and Education of Spain, the Science and Technology Facilities Council of the United Kingdom, the Higher Education Funding Council for England, the National Center for Supercomputing Applications at the University of Illinois at Urbana-Champaign, the Kavli Institute of Cosmological Physics at the University of Chicago, Center for Cosmology and Astro-Particle Physics at the Ohio State University, the Mitchell Institute for Fundamental Physics and Astronomy at Texas A\&M University, Financiadora de Estudos e Projetos, Fundacao Carlos Chagas Filho de Amparo, Financiadora de Estudos e Projetos, Fundacao Carlos Chagas Filho de Amparo a Pesquisa do Estado do Rio de Janeiro, Conselho Nacional de Desenvolvimento Cientifico e Tecnologico and the Ministerio da Ciencia, Tecnologia e Inovacao, the Deutsche Forschungsgemeinschaft and the Collaborating Institutions in the Dark Energy Survey. The Collaborating Institutions are Argonne National Laboratory, the University of California at Santa Cruz, the University of Cambridge, Centro de Investigaciones Energeticas, Medioambientales y Tecnologicas-Madrid, the University of Chicago, University College London, the DES-Brazil Consortium, the University of Edinburgh, the Eidgenossische Technische Hochschule (ETH) Zurich, Fermi National Accelerator Laboratory, the University of Illinois at Urbana-Champaign, the Institut de Ciencies de l’Espai (IEEC/CSIC), the Institut de Fisica d’Altes Energies, Lawrence Berkeley National Laboratory, the Ludwig Maximilians Universitat Munchen and the associated Excellence Cluster Universe, the University of Michigan, NSF’s NOIRLab, the University of Nottingham, the Ohio State University, the University of Pennsylvania, the University of Portsmouth, SLAC National Accelerator Laboratory, Stanford University, the University of Sussex, and Texas A\&M University.

BASS is a key project of the Telescope Access Program (TAP), which has been funded by the National Astronomical Observatories of China, the Chinese Academy of Sciences (the Strategic Priority Research Program “The Emergence of Cosmological Structures” Grant \# XDB09000000), and the Special Fund for Astronomy from the Ministry of Finance. The BASS is also supported by the External Cooperation Program of Chinese Academy of Sciences (Grant \# 114A11KYSB20160057), and Chinese National Natural Science Foundation (Grant \# 11433005).

The Legacy Survey team makes use of data products from the Near-Earth Object Wide-field Infrared Survey Explorer (NEOWISE), which is a project of the Jet Propulsion Laboratory/California Institute of Technology. NEOWISE is funded by the National Aeronautics and Space Administration.

The Legacy Surveys imaging of the DESI footprint is supported by the Director, Office of Science, Office of High Energy Physics of the U.S. Department of Energy under Contract No. DE-AC02-05CH1123, by the National Energy Research Scientific Computing Center, a DOE Office of Science User Facility under the same contract; and by the U.S. National Science Foundation, Division of Astronomical Sciences under Contract No. AST-0950945 to NOAO.
\end{acknowledgments}


\bibliography{sample631}{}
\bibliographystyle{aasjournal}



\end{document}